*Optical and X-ray Photo-emission Spectroscopies of Core/Shell Colloidal CdSe/CdS Quantum Dots: Modeling and Experimental Determination of Band Alignment*

*Damien Simonot[1], Céline Roux-Byl[2], Xiangzhen Xu,[2] Willy Daney de Marcillac[1], Corentin Dabard[1,2], Mathieu G. Silly[3], Emmanuel Lhuillier[1], Thomas Pons[2], Simon Huppert[1], Agnès Maître[1, *]*

[1] *Institut des Nanosciences de Paris, Sorbonne Université, CNRS UMR 7588, 4 place Jussieu, 75252 Paris Cedex 05, France*
[2] *Laboratoire de Physique et d'Etude des Matériaux, ESPCI-Paris, PSL Research University, Sorbonne University, CNRS UMR 8213, 10 rue Vauquelin 75005 Paris, France*
[3] *Synchrotron SOLEIL, L'Orme des Merisiers, Départementale 128, 91190 Saint-Aubin, France.*

[*] Contact author: agnes.maitre@insp.jussieu.fr

**Abstract**

Optical properties of multilayer semi-conductor nano-emitters are crucially dependent on the relative energy levels of their different components. For core/shell quantum dots, the relative energy difference between conduction band edge of core and shell materials induces, depending on its value, either a confinement of the electron within the core or a delocalization of its wave function within the whole quantum dot. This results in drastic consequences on the energy and the oscillator strength of the fundamental transition. Surprisingly, the literature currently lacks a definitive value for the energy difference between CdSe and CdS conduction band edges as most of the experimental studies provide values corresponding to specific geometries of quantum dots. Here, we develop a full theoretical model expressing energy levels considering core/shell interface pressure, ligands and enabling the accurate prediction of the bandgap value with the nanocrystal size. It allows to reliably determine the energy difference between the conduction band edge of CdSe and CdS materials, known as the conduction band offset, in such a way that this value can later be used to model quantum dots of any geometry. This value is determined using our model and two different experimental methods: optical spectroscopy and X-ray photoemission (XPS) experiments.

CdSe/CdS core/shell nanocrystals are exceptional nano-emitters that exhibit intriguing emission characteristics at both the individual and collective levels. These nanocrystals serve as efficient single-photon sources at room temperature, offering high quantum efficiency up to 99.5% [1,2]. They find a wide range of applications in optoelectronics [3,4] and biophotonics [5,6]. These quantum dots can be employed as fundamental building blocks for laser sources integrated into nanophotonic devices or utilized as biomarkers in microscopy [7]. Consequently, controlling and predicting the luminescence mechanisms in these structures is essential. The synthesis of CdSe/CdS core/shell heterostructures enables customizing their optical properties in emission and absorption by adapting the size of the core and the shell [8]. Indeed, one of the key ingredients in the luminescence of such emitters is the relative energy difference, also called conduction band offset, between the conduction band of CdS and CdSe. It plays a significant role in determining the degree of delocalization of the electron in the shell, and thus the emission properties of nanocrystals. However, its precise value remains surprisingly vague in the literature ranging from -0.3 to 0.4 eV [9–12]. Scanning tunneling spectroscopy experiments on nanorods provided a value of 0.2 eV [13], while intra-band measurements determined the offset $O_c$ to be close to 0.18 eV [14]. In this paper, we will determine a reliable value for the CdSe/ CdS conduction band offset by developing a comprehensive theoretical model. It accounts accurately for quantum confinement, Coulomb electron-hole interaction, as well as for the effect of mechanical strain on the energy bands. This latter point proves crucial to correctly predict optical properties of core/shell CdSe/CdS QDs, and experimental studies on absorption transitions.

Indeed, the optical absorption mechanism is generally characterized by the exciton wave function, the transition's oscillator strength, and the associated energy. Determining all of them is a challenge that the scientific community has actively worked on. Atomic-scale electronic structure calculations such as with the density functional theory (DFT) can in principle provide optical transition energies [10,15,16] but their numerical cost for nanometric systems such as CdSe/CdS quantum dots can be very high (though empirical or semi-empirical approximations have been proposed to reduce the computational load [17,18] ). Furthermore, the influence of the physical and structural parameters of the quantum dots on its optical properties is often not straightforward to interpret in all-atoms simulations. Numerous studies therefore rely on the effective mass approximation (EMA), a simpler yet accurate framework for electronic structure calculations [19] . In particular, pioneering works based on an 8-bands EMA model provided predictions of the excitonic fine structure for spherical nanocrystal, including electron-hole exchange interaction [17,20,21] . Other works have employed a two-bands description, in order to obtain a simpler and more versatile framework [22–24] , in which additional effects can be included such as a treatment of Coulomb interaction beyond perturbation theory [25,26] , non-standard band alignment configurations [27] or the modelling of Auger processes [28].

Simultaneously, various studies have explored the impact of strain [29,30] and temperature [31] on the energy bands of semiconductors. They have demonstrated band shifts under pressure, dependent on dimensions [32]. Organic ligands grafted at the surface of colloidal nanocrystals in order to prevent aggregation, play a central role on many aspects of QD properties, including transport in films [33] and surface passivation [34]. They are also important in determining the boundary conditions for the electron and hole wavefunctions at the nanocrystal surface and can be considered as an external high bandgap semiconductor layer with electron and hole effective masses to be determined. Indeed, without considering the ligand layers, confinement models of electron and hole in QDs do not provide an accurate prediction of CdSe QD bandgap variations with size, a problem long considered as "swept under the rug" [35] .

Most existing studies for transition energy calculations address these effects independently. In this paper, we will highlight these previous works by proposing a new method providing a more integrated and detailed approach. Firstly, the fundamental excitonic transition is modeled with a comprehensive theoretical description that accounts for electronic confinement (including the influence of the surrounding ligands), Coulomb electron-hole interaction, mechanical strain effects at the core/shell interface as well as an effective correction term to approximate exchange interactions. This theoretical study is associated to two complementary experiments, enabling the determination of the offset between the CdSe and CdS conduction bands. We first analyze the experiment based on absorption spectroscopy on CdSe/CdS quantum dots, before discussing X-ray photoemission on large CdSe and CdS spherical nanocrystals.

## I. THEORETICAL MODEL
### A. Electronic transition for core/shell/ligands quantum dots

In this work, CdSe and CdS quantum dots are synthetized both in the wurtzite structure characterized by a single conduction band and three valence bands: heavy hole (hh), light hole (lh) and split-off (SO). The latter is separated from the two others by a relatively large splitting equal to 0.42 eV and will be disregarded in the following. For bulk materials, the gap between the conduction band and the heavy-hole valence band, $E^g$, has a value of 1.75 eV and 2.50 eV for the CdSe and CdS respectively [36] . The light hole band is separated from the heavy hole valence band by a crystal-field induced splitting $\Delta_{cr}$ amounting to 0.025 eV for CdSe and 0.040 eV for CdS [36] (Fig. 1).

Each quantum dot is composed of a CdSe core with a diameter denoted as $d_1$, surrounded by a protective CdS shell with a thickness $t_2$. This protective shell serves to prevent blinking and to increase emission intensity [37,38]. Consequently, the total diameter of the quantum dot can be expressed as $d_{tot} = d_1 + 2t_2$. We consider that the quantum dot has a spherical symmetry.

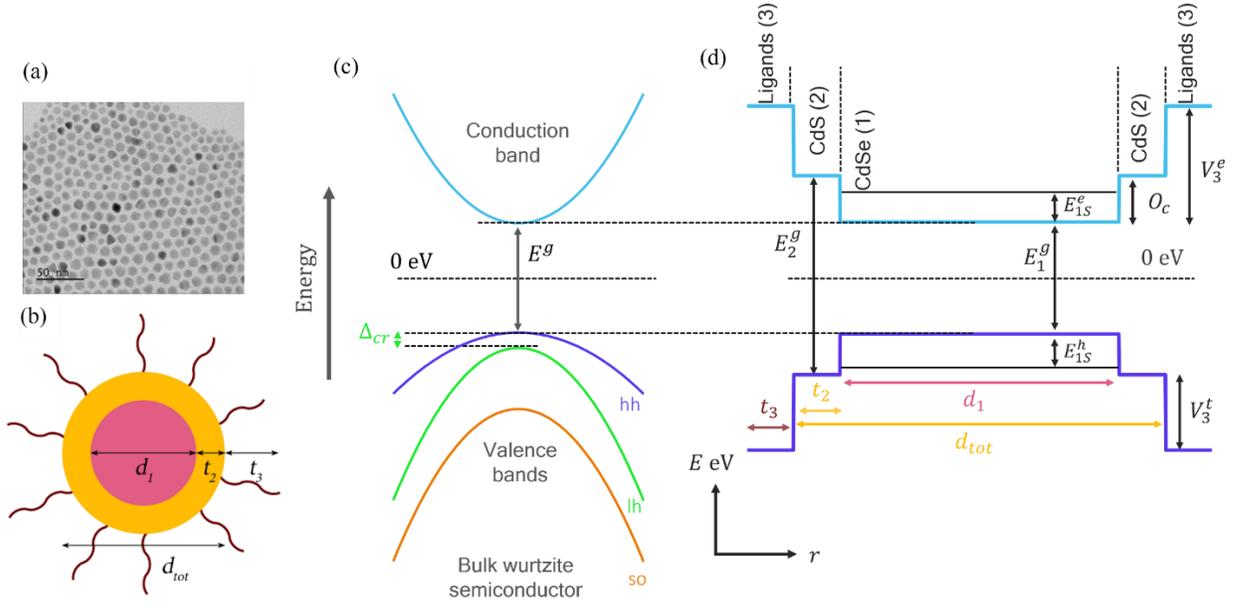

*FIG. 1. (a) Transmission Electron Microscopy (TEM) visualization of spherical CdSe/CdS nanocrystals. (b) Scheme of a spherical CdSe/CdS core/shell nanocrystal with ligands assuring colloidal stability and improving fluorescence intensity. (c)Energy band diagram for wurtzite bulk CdS or CdSe. There is one conduction band and three valence bands: heavy-hole (hh), light-hole (lh) and split-off (so). The gap between the conduction band and the heavy hole valence band is equal to $E^g$. The split between the heavy- and light-hole valence bands is equal to $\Delta_{cr}$. (d) Spatial representation of electron and heavy-hole confinement potentials for CdSe/CdS "bulk" quantum dots: In this representation the colored energy levels correspond to the energy levels of the bulk materials and no finite size effect of the quantum dot is considered in this figure. The ligands are approximated as a semiconductor shell of thickness $t_3$. $E_1^g$ and $E_2^g$ are the bulk gap energy of respectively the CdSe and CdS. $E_{1S}^e$ ($E_{1S}^h$) is the fundamental confined energy of the electron (hole).*

Determining the exciton energy and characteristics in such structures requires solving the steady state Schrödinger equation for the coupled electron and hole in the effective mass approximation. This poses significant analytical and numerical challenges [26]. Here we make use of two main approximations: first we assume that the (lh) and (hh) hole states can be represented as decoupled parabolic bands, second, we make use of the Hartree approximation [25,26] for the electron-hole interaction. This allows representing the exciton wave function as the product of the electron wave function $\Psi_{n,l,m}^e$ and the hole wave function $\Psi_{n',l',m'}^h$. $e$ and $h$ denotes respectively the electron and the hole. $n, l, m, n', l', m'$ are the quantum numbers associated to the spherical symmetry. To simplify the notation, we define $\Psi_{n,l,m}^{\{e,h\}} = \{\Psi_{n,l,m}^e, \Psi_{n',l',m'}^h\}$, then we can write:

$$\Psi_{n,l,m}^{\{e,h\}} = \frac{R^{\{e,h\}}(r)}{r} Y_{n,l,m}(\theta, \phi) \qquad (1)$$

$r$ is the radial coordinate; $\theta, \phi$ are the polar and azimuthal angles respectively. $Y_{n,l,m}(\theta, \phi)$ are the spherical harmonics and $R^{\{e,h\}}(r)$ are the radial parts of the wave functions.

In this paper, we consider transitions close to the band edge and thus we exclusively focus on the values of the fundamental excitonic transition where $n = n' = 1$, and $l = l' = m = m' = 0$. Consequently, the electron and hole are both in the 1S state. In the case of the 1S1S transition, $Y(\theta, \phi) = \frac{1}{\sqrt{4\pi}}$.

Consequently, the coupled Schrödinger equations of the 1S1S exciton are written as:

$$\begin{cases} \left[-\frac{\hbar^2}{2}\frac{1}{m^e(r)}\frac{d^2}{dr^2} - \frac{1}{2}e\phi^h(r) + V^e(r) + \Delta E^e(r)\right] R_e(r) = E^e R^e(r) \\ \left[-\frac{\hbar^2}{2}\frac{1}{m^h(r)}\frac{d^2}{dr^2} - \frac{1}{2}e\phi^e(r) + V^h(r) + \Delta E^h(r)\right] R_h(r) = E^h R^h(r) \end{cases} \quad (2)$$

$\hbar$ represents the reduced Planck constant, $m^{\{e,h\}}(r)$ denote the effective masses of the electron and the hole. In both equations, the first term corresponds to kinetic energy, while the second term accounts for the Coulombic interaction between the electron and the hole. $\phi^{\{e,h\}}(r)$ is the electrostatic potential produced by the electron (hole), acting on the hole (electron). The inclusion of the 1/2 factor is intended to prevent redundant consideration of the attractive Coulomb potential. $\Delta E^{(e,h)}(r)$ is the adjustment of the potential due to other factors like mechanical strain (see in the following). The electrostatic component is computed by solving the Poisson equation. For a spherically symmetric charge density, the Poisson equation reads:

$$\frac{\varepsilon_0}{r}\frac{d^2}{dr^2}\left[r\kappa_1\phi^{\{e,h\}}(r)\right] = \pm e\rho^{\{e,h\}}(r) \quad (3)$$

$\varepsilon_0$ is the permittivity of vacuum. As an approximation, we will consider the permittivity $\varepsilon_0\kappa_1$ of the nanocrystal to be constant and equal to the CdSe one (See SI). $e\rho^{\{e,h\}}(r)$ is the charge density created by an electron or a hole.

$$\rho^{\{e,h\}}(r) = \left|\Psi_n^{\{e,h\}}\right|^2 \quad (4)$$

To guarantee a correct representation of the electrostatic potential (that effectively spreads beyond the limits of the nanocrystal), $\phi^{\{e,h\}}$ are computed over a domain extending to $r = 20$ nm, which we numerically checked to be sufficient. It is important to note that the Coulomb potential represents an attractive force between the electron and the hole, which results in the reduction of the exciton's energy. $V^{\{e,h\}}(r)$ are the energy potentials of the electron and hole defined by the conduction and valence bands of the core, shell and ligands regions. The energy levels of the conduction bands in both the core and shell are established based on the electron affinity to the vacuum within the bulk materials. The relative difference between the confinement potentials of core and the shell conduction bands is called conduction offset $O_c = V^e_{CdS} - V^e_{CdSe}$ (Fig. 1(d)).

The Schrödinger electron and hole equations (2) and the Poisson equation (3) are solved by discretizing the radial variable $r$ on a grid, using finite difference formulas for the spatial derivatives. Since (2) and (3) are mutually dependent on each other, they are solved together in an iterative self-consistent scheme [26] (details in SI). The exciton energy is finally obtained as $E_{exc} = E^e_{1S} + E^h_{1S}$

For colloidal quantum dots, organic ligands are grafted all around the shell ensuring not only colloidal stability but also improving the fluorescence through passivation of the surface dangling bond. Indeed, they play a crucial role in the realistic modeling of the quantum dot as they allow the spreading of the electrons and holes wave functions beyond the core/shell sphere. Consequently, to render this effect, a third semiconductor shell layer is introduced to describe the ligands surrounding the quantum dot. This layer has a thickness $t_3 = 4$ nm, and is modeled with effective mass parameters, namely $m^e_3$ for the electron and $m^h_3$ for the hole, along with conduction and valence band barriers, $V^e_3$ and $V^h_3$, respectively (Fig. 1(d)).

Finally, the potentials $V^e(r)$ and $V^h(r)$ are expressed as:

|  | $r \in$ core (1) | $r \in$ shell (2) | $r \in$ ligands (3) |
|---|---|---|---|
| $V^e(r)$ | $\frac{1}{2}E^g_1$ | $\frac{1}{2}E^g_1 + O_c$ | $\frac{1}{2}E^g_1 + V^e_3$ |
| $V^h(r)$ | $-\frac{1}{2}E^g_1$ | $\frac{1}{2}E^g_1 + O_c - E^g_2$ | $-\frac{1}{2}E^g_1 - V^h_3$ |

Finally, important works [20] by Efros and coworkers have shown that the electron-hole exchange interaction can significantly affect the exciton energy in CdSe quantum dots of small diameter sizes. Since our model does not explicitly account for exchange effects, we added the corresponding correction a posteriori to our results for the exciton energy (see details in SI).

## B. Stress induced energy shifts

For a more realistic description, it is imperative to consider the stress induced by the shell on the core by epitaxial growth (Fig. 2(a)). Indeed, there is a relative lattice mismatch of approximately 3.8% between CdSe and CdS [39]. The core lattice constant, denoted as $a_1$, is larger than the shell lattice constant, denoted as $a_2$. Consequently, this results in a radial displacement of the atoms, in opposite direction for core and shell ones, with respect to the unconstrained configuration. This results in a compression of the core and a dilatation of the shell (Fig. 2(a)). We neglect any pressure which could be due to organic ligands fixed at shell outer interface [40]. To ascertain the radial displacements associated with the core/shell pressure, a spherically symmetric elastic continuum model provides an expression for the radial displacement of a hollow sphere with inner and outer radius subject to both inner and outer pressures [30,41].

The pressure at the interface, denoted as $P_{int}$, is determined from the elastic sphere model. It is contingent upon the material properties specifically the Poisson ratio ($\nu_1$ and $\nu_2$) and the Young modulus ($Y_1$ and $Y_2$) of both the core and shell (details in SI). It also varies with the dimensions of the core diameter and the thickness of the shell (see Fig. 2). Within the core, the pressure $P_{int}$ remains constant. It tends to rise as the shell thickness increases (Fig. 2(b)) and/or as the core diameter decreases (Fig. 2(c)). In the shell, the pressure is maximum at the core/shell interface and minimum at the shell outer interface, tending to 0 when the shell thickness becomes very large.

The pressure effects described here result in a shift $\Delta E = E_{pressure} - E_{without\ pressure}$ of the energy bands. These shifts [32] $\Delta E_{\{1,2\}}^{\{e,h\}}$ of the conduction (e) and valence (h) bands for both core (1) and shell (2) can be calculated using the deformation potential $\alpha$, which is expressed as [42]:

$$\alpha = \frac{\partial E}{\partial (\ln V)} \quad (6)$$

Where $E$ represents the energy, and $V$ represents the volume. It is important to note that the deformation potential of the conduction band $\alpha^e$ is negative whereas the valence one $\alpha^h$ is positive. Moreover, the absolute value of $\alpha^e$ is larger than $\alpha^h$. Following Jing and al [32] the conduction and valence band shift for both the core and shell can be expressed as (details in SI):

$$\Delta E_1^{\{e,h\}} = 3\alpha_1^{\{e,h\}} P_{int} \frac{2\nu_1 - 1}{Y_1}$$

$$\Delta E_2^{\{e,h\}}(r) = 3\alpha_2^{\{e,h\}} \frac{P_{int}}{Y_2} \frac{1}{\left(\frac{r_2}{r_1}\right)^3 - 1} \left((1 - 2\nu_2) + \frac{(1+\nu_2)}{2}\left(\frac{r_2}{r}\right)^3\right) \quad (7)$$

The energy shift of the conduction band $\Delta E_{\{1,2\}}^e$ as a function of the position $r$, for a fixed core radius and an increasing number of CdS layers, is shown in Fig. 2(d). Similarly, the variation of the conduction band $\Delta E_{\{1,2\}}^e$ as a function of the position $r$, for a fixed shell thickness and an increasing core radius, is shown in Fig. 2(e). Both Figs. 2(d) and 2(e) illustrate that the potential confinement $V^{\{e,h\}}(r)$ is dependent on both the core diameter and the thickness of the shell.

Within the core, as the pressure is constant and positive, the energy shifts are constant across radial positions, positive for conduction band and negative for valence band. This leads to an increase in energy of the core bandgap.

Within the shell, the energy shift is negative in the conduction band. As the pressure is diminishing with the radial position, its absolute value is maximum at the core/shell interface, reducing with $r$. At the shell outer interface, the energy tends towards that obtained without considering the pressure, especially for thick shell.

At the shell/core interface, the energy shift undergoes an abrupt change, from positive to negative values for the conduction band. Therefore, this abrupt change due to mechanical strain tends to change significantly the electron energy barrier at the core /shell interface. Consequently, the resulting pressure induced by the shell, smoothes the energy profile and lets the electron wavefunction spread further out of the core (see Fig. 2(f)). Such unsharp energy profile has generated a lot of interest over the recent years to reduce Auger relaxation [43] that is a major non radiative recombination pathway for colloidal nanocrystals. Its non-radiative fast relaxation challenges the design of lasers based on quantum dots [28] . In practice, the most investigated approach to design such non sharp energy profile has been focused on graded composition, which increases the complexity of the growth process [44] . From our simulation, it appears that pressure generates a similar effect, suggesting that lattice mismatch may also have beneficial aspect.

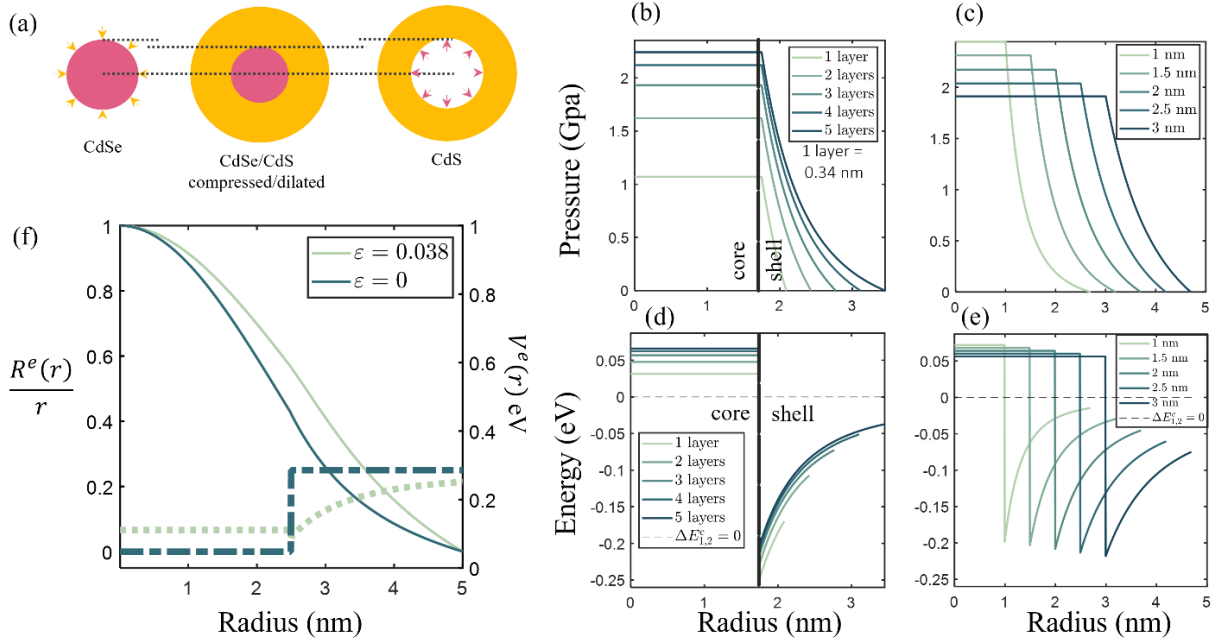

*FIG. 2. (a) The CdSe core is compressed by the CdS shell meanwhile the shell is dilated. (b-d) CdSe core radius equal to 1.75 nm, with 1 to 5 shell layers. One single CdS layer has a thickness of 0.34 nm. (c-e) The number of CdS shell layers is fixed to 5. The CdSe core radius varies between 1 and 3 nm. **b-c)** Pressure within a CdSe/CdS core/shell quantum dot as a function of the radial position. (d-e) Evolution of the conduction band of the core/shell quantum dot ΔE in function of the position r. The dashed lined represent the conduction band without taking into account the pressure and considering a conduction offset $O_c$ = 0 eV. (f) Representation of the normalized wave function of the electron in the ground state and the confinement potential of the corresponding conduction band considering an offset of 0.25 eV. ε represents the lattice mismatch between CdSe and CdS; the case ε = 0 means that we do not consider the effects of pressure.*

To summarize, our model for the calculation of exciton levels in core/shell/ligands quantum dot includes several crucial correcting factors, such as Coulomb interaction, presence of ligands, mechanical strain at the core/shell interface and approximates exchange effects. This comprehensive model enables the determination of the energy associated with the fundamental $1S1S_{hh}$ transition, given an offset $O_c$ between conduction band of bulk CdSe and CdS.

We will now use optical spectroscopy and XPS measurements, to extract the offset $O_c$ value between bulk CdSe and CdS from experimental data and preceding model.

## II. Results and discussion
### A. Optical spectroscopy

Most of the parameters in the model presented above are known from the literature, but it is not the case for the parameters to apply to the ligand layer in the context of an effective semiconductor layer description. We thus start by determining these ligands parameters (effective masses and energy levels) by modeling the simple case of pure CdSe quantum dots.

Six samples of CdSe core-only quantum dots with different diameters, ranging from 3.2 nm to 5.9 nm, have been synthesized using a standard solvothermal approach. [45] They are diluted in hexane and are surrounded by oleate and oleylamine organic ligands. The collective absorption spectrum is measured at room temperature for each solution (Fig. 3(a)). From the minima of the second derivative of the experimental absorption spectra, the energy transitions 1S1S$_{hh}$ of the heavy hole and transition 1S1S$_{lh}$ of the light hole are determined (Fig. 3(b)) [46]. These quantum dots do not have any CdS shell, so that neither the offset nor the pressure have to be taken into account in the modelling. Considering electron and hole wave functions, ligands effective masses and energy levels have indeed a correlated influence on the wave function spreading. Fixing electron and hole barriers to the literature value of 0.5 eV above the CdSe conduction band [33] (or below the valence bands for holes), leads to the determination of the ligand's electron and hole effective masses $m_3^e$ and $m_3^h$. This leads to a good agreement between theoretical predictions and experimental data as shown in Fig. 3(c), with effective masses of the ligands being $m_3^{\{e,h\}} = 0.025\,m_0, 1.48\,m_0$. Note that this description of the ligand layer is approximate and only meant to capture the slight delocalization of the electrons outside of the nanocrystal boundaries, but the effective ligands parameters extracted by this procedure might not have a direct physical interpretation. As shown in Fig. 3(c) our model provides an excellent prediction of the bandgap values as a function of the core size.

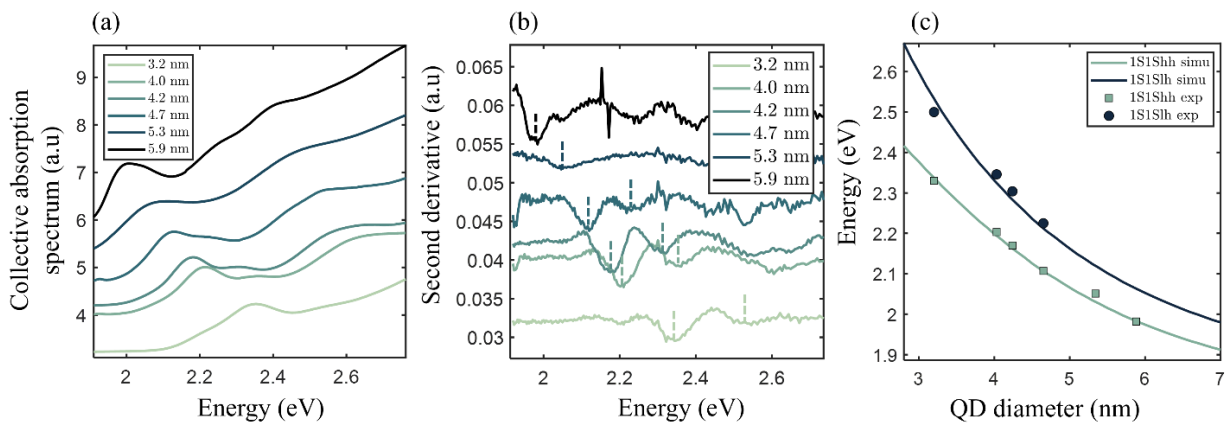

*FIG. 3. (a) Absorption spectra of CdSe quantum dots solution for different diameters (b) second derivative of the absorption spectra (c) Experimental (points) and theoretical (lines) evolution of the 1S1S$_{hh}$ and 1S1S$_{lh}$ transition of CdSe/ ligands nanocrystal as a function of the diameter.*

In a second step, we aim at modeling the optical transitions in CdSe/CdS core/shell QDs and evaluating the offset $O_c$ between the conduction bands of CdS and CdSe. We consider core/shell CdSe/CdS/ligands quantum dots diluted in hexane solutions prepared following a layer-by-layer approach [47]. From 2 different seeds of CdSe, the thickness of the CdS layer is controlled from one to six CdS monolayers. The first seed has a core diameter $d_1 = 4.0$ nm. while the second seed one is $d_1 = 5.9$ nm. We obtain 13 solutions with different core diameters and shell thicknesses and perform spectroscopic measurements at room temperature. As above, using the second derivative of absorption spectra, we determine the experimental 1S1S$_{hh}$ transition energies. The experimental evolution of the fundamental exciton level for both series of samples as a function of the total diameter (and the number of CdS layers) is plotted in Fig. 4. The squares represent the experimental data, and the green dotted line the simulation.

For the smallest quantum dot having no CdS shell, the absorption wavelength is shorter because both the electron and the hole are confined only in the core. For core/shell quantum dots, a smaller core diameter induces a stronger exciton confinement, leading to a shorter absorption wavelength. For a given core size, when CdS layers are added, the wave functions spread in the shell, particularly for the electron, resulting in a weaker confinement energy and therefore a longer corresponding absorption wavelength. After four CdS layers, the electron confinement starts to become negligible and the variation of the absorption wavelength begins to saturate.

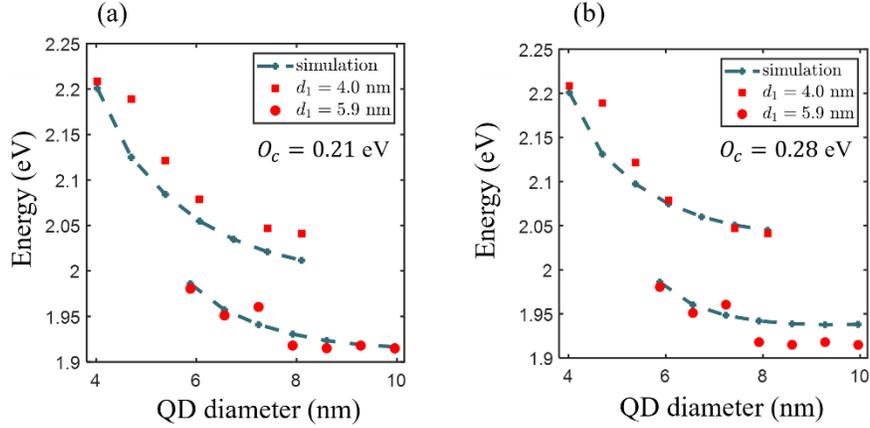

**FIG. 4.** *Wavelength of the fundamental transition of core/shell nanocrystal as a function of the QD diameter. From different CdSe cores of 4nm and 6 nm, 1 to 6 monolayers of CdS are added. Squares: Experimental data. Line: fit of experimental data with our model (a) The fit is performed on the thick core QD (data in back) giving an conduction offset $O_c$ of 0.21 eV. (b) The fit is performed on the small core QD (data in red) giving an offset $O_c$ of 0.28 eV.*

Finally, we use our theoretical model to fit the experimental fundamental transition as a function of the quantum dot diameter by our theoretical model. We calculate and minimize with the offset $O_c$ as fitting parameter, the function $L[O_c] = \frac{1}{N}\sum \left|\frac{E_{exp}-E_{theo(o_c)}}{E_{exp}}\right|$, corresponding to the normalized distance between the experimental energy $E_{exp}$ and theoretical one $E_{theo}(O_c)$ (see SI). When the fit is performed on the data corresponding to the bigger core (diameter = 5.9 nm) the best fit offset value is 0.21eV ± 0.05 eV. When it is performed on the smaller core (diameter = 4 nm) it is found to be equal to 0.28 eV ± 0.04 eV.

A slightly higher offset values for small cores may be due to effects not considered in our model. A recent publication [48] underlines a strain induced piezo electric effect which results in a higher confinement of electrons within the core and holes within the shell, leading to a higher offset for larger strain, as it is the case for smaller cores.

Nonetheless, both offset values are in accordance, setting a reliable value of an CdSe/CdS offset $O_c$ of 0.2-0.3 eV. It leads to a best value for the offset $O_c = 0.25$ eV, a value that should be used in future modelization to provide a good description of CdSe/CdS quantum dots of all possible size and geometry.

To confirm these results, we perform in the following a complementary experiment based on X-ray Photoelectron Spectroscopy (XPS) in order to determine the offset by another method. In absorption spectroscopy, we have determined an offset using quantum dots with an CdSe/CdS interface, whereas in XPS, we will perform two successive experiments with core only QDs, one with CdSe QDs and the other one with CdS. By comparing conduction offsets values by both methods, we are able to point out a possible physical effect at the interface not taken into account in our model.

**B. XPS experiment**

X-ray Photoelectron Spectroscopy (XPS) measurements rely on the interaction between X-rays and the surface of a sample. This interaction results in the ejection of electrons from the outermost layer of the sample, which are subsequently analyzed based on their kinetic energy. When dealing with a semiconductor sample, this analysis provides insights on the position of the Fermi level $E_f$ considering the vacuum energy $W_v$ set to 0 eV. The energy level of the valence band $W_v$ relative to the Fermi level [49] are also determined. These levels, $E_f$ and $W_v$, are visually depicted in Fig. 5(a).

We use two sets of samples. The first set comprises a layer of CdSe nanocrystals with a large diameter of 7.2 nm, while the second set features a layer of CdS nanocrystals with a larger diameter of 10.9 nm. The goal was to work with nanocrystals large enough to minimize confinement effects on energy levels. As a result, the conduction band positions of these samples should be closely aligned with the conduction bands of their respective bulk materials.

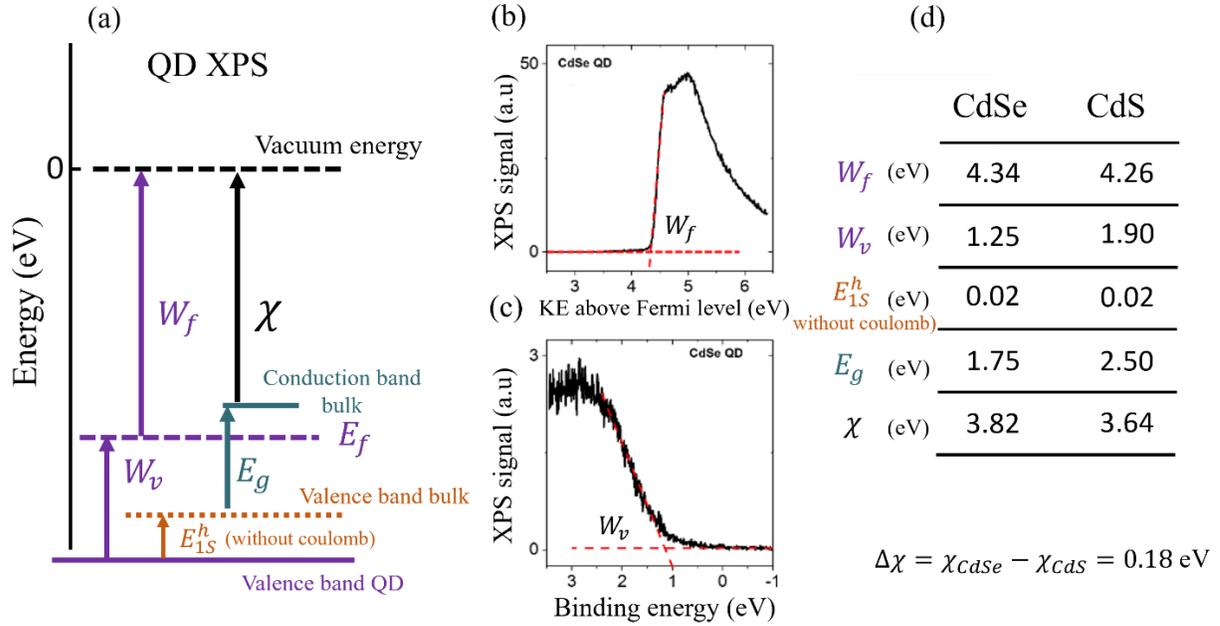

**FIG. 5.** (a). Energy level scheme for quantum dots. $W_v$ and $W_f$ are extracted by XPS absorbance spectrum. $W_f$ is the energy difference between vacuum and Fermi level. $W_v$ is the energy difference between QD valence band and Fermi level. $E_{1S}^h$ is the hole confinement energy for CdSe and CDS QD (diameter 8.9 nm and 10.9 nm respectively), $E_g$ is the bulk gap energy for the bulk material. $\chi$ the energy difference between bulk conduction band and vacuum level is the electronic affinity. (b) X-ray photoemission spectrum relative to the cut off of the secondary electron of the CdSe QD (diameter 8.2 nm) (c). X-ray photoemission spectrum relative to valence band of the same CdSe QD. Spectra for CdS are represented in SI. (d) Physical numerical values for CdS and CdSe energy. violet: XPS; brown: simulation; green: literature; black: electronic affinity calculated from previous values for CdS

We conducted XPS experiments at synchrotron SOLEIL. The measurement results are shown Figs. 5(c) and 5(d) and summarized in Fig. 5(b). First, the cut off measurement of the secondary electrons of the CdSe QD (Figs. 5(b) and S1(b) resp. for CdS) sets the Fermi level 4.34 eV below vacuum (resp. 4.26 eV for CdS). In a second time, the valence band energy relatively to the Fermi level $W_v = E_v - E_f$ is measured at 1.25 eV for CdSe quantum dots (Fig. 5(c), resp. 1.90 eV for CdS in SI Fig. S1(a)).

Finally, we want to determine, for bulk materials, the valence band energy. The confinement energy, which have been calculated using our model (0.02 eV, for our CdSe and CdS quantum dots), are subtracted from the measured valence band energy for large quantum dots. It leads via the tabulated bulk band gap energy, to the determination of the conduction band edge energy for bulk CdSe and CdS relative to the vacuum energy, corresponding to the electronic affinity for both materials. The difference $\Delta\chi$ correspond to the offset value $O_c$ and is estimated to be equal to 0.18 eV. This value is in very good accordance with the one obtained using optical spectroscopy on CdSe/CdS quantum dots.

### C. XPS/ optical spectroscopy experiment

Both experiments give similar offset values, slightly larger for optical spectroscopy. The small difference $\delta O_c$ between both experiments, between 20 meV for large CdSe/CdS quantum dots, and 100 meV for small quantum dots can be reminiscent of the strain induced piezoelectric effect [48] .Moreover, in our model used for optical spectroscopy fitting, we assume for CdSe/CdS QD, a sharp transition in composition between CdSe and CdS whereas the real transition is smoother. Indeed an initial partial dissolution of the CdSe core is frequently observed at early stages of the synthesis due to heating in the presence of an excess of oleylamine ligands [50] . It leads to an effective slightly smaller core, which explains the slightly different effective offset values for different core sizes.

In summary, a model has been proposed to determine the lowest optical transition energy, and the corresponding wave function, of an exciton in a spherical core-shell quantum dot where ligands are considered as an additional semiconductor layer. This comprehensive model leads to a very good prediction of fundamental transition energies as a function of the CdSe quantum dot size. To determine the band alignment between CdSe and CdS, from two cores of different diameter, one to six CdS layers are added, and the energy of the fundamental transition is measured from the absorption spectrum. By fitting the experimental data with the model, we determined a conduction offset value $O_c$ ranging from 0.2 to 0.3 eV. These measurements have been confirmed by XPS experiments. Our model and both experiments make it possible to determine an offset value for bulk CdSe and CdS which can be used to determine the fundamental transition of any core/shell/ligands CdSe/CdS quantum dots. These protocols and model can be extended to more complex nanostructures, such as nanocrystals composed of core/shell/shell/ligands and to a wide range of material, opening the way to a more comprehensive prediction of spectroscopic properties of semiconductors.

## ACKNOWLEDGEMENTS

The authors thanks Robson Ferreira, Francesca Carosella and Debora Pierucci for fruitful discussions.

The project is supported by ERC AQDtive (grant n°101086358). This work was supported by French state funds managed by the Agence Nationale de la recherche (ANR) through the grant E-map (ANR-22-CE50-0025) and CoLiME (ANR-23-CE09-0029)